\documentclass[12pt]{article}
\usepackage{amsxtra,amssymb,amsthm,amsmath,latexsym}

\theoremstyle{plain}

\def\oH{\buildrel\circ\over H}
\def\oH1{\buildrel\circ\over H\kern-.02in{}^1}
\def\oH1{\buildrel\circ\over H\kern-.02in{}^1}

\def\d{\delta}

\begin{document}    

Nonlinear Funct. Anal. Appl., 8, N2, (2003).

\title{
 On the discrepancy principle
}
\author{
A.G. Ramm\\
Mathematics Department, Kansas State University, \\
 Manhattan, KS 66506-2602, USA\\
ramm@math.ksu.edu\\ 
}

\date{}
\maketitle\thispagestyle{empty}

\centerline{ Abstract}
A simple proof of the convergence of the variational regularization,
with the regularization parameter, chosen by the discrepancy 
principle, is given
for linear operators under suitable assumptions.
It is shown that the discrepancy principle, in general, does  not yield 
uniform with respect to the data convergence. An a priori choice of the 
regularization parameter is proposed and justified for some nonlinear
operator equations.

\section{Introduction. }

In this note the convergence of the variational regularization
with the regularization parameter chosen by the discrepancy 
principle is discussed for linear operators.
In [1] and [2] one finds the references related to the discrepancy 
principle.  

Our assumptions in this Section are:

i) {\it $A$ is a linear, bounded, injective 
operator on a Hilbert
space $H$,  the equation $A(y)=f$
has a solution $y$, $A^{-1}$ is not continuous, so that (1.1) is an 
ill-posed problem, $||f_\d -f||\leq \d$, $||f_\d||> \d$. 
}

Given $\{\d,f_\d\}$, one wants to approximate $y$ stably in the sense
$$\lim_{\d\to 0}||R_\d f_\d-y||=0, \eqno{(1.1)}$$
 where $R_\d$ is an operator defined on all of $H$, a regularizer. Let 
$F(u):=||Au-f_\d||^2+a||u||^2$, where $a>0$ is a parameter, and let $u_a$
be the solution to $F(u_a)=\inf F(u)$. Existence and uniqueness of this
variational problem are known: if $B:=A^*A$, then
$$u_a=(B+a)^{-1}A^*f_\d,\quad (B+a)u_a=A^*f_\d, \eqno{(1.2)}$$
equation (1.2) is the Euler equation for $F$, it has a unique solution
$u_a$ since $B\geq 0$ and $a>0$, and 
$$
F(u_a)\leq F(u), \quad ||Au_a-f_\d||^2 +a||u_a||^2\leq \d^2+a||y||^2.
 \eqno{(1.3)}$$ 
Assume $f_\d\perp N(A^*)$, where $N(A^*)$ is the null-space of $A^*$.
This is done without loss of generality, since $f_\d$ enters under the 
sign of $A^*$ in (1.2). The discrepancy principle (DP) (introduced
by Morozov, see [1] and [2]) chooses 
$a=a(\d)$ as the root of the equation:
$$
h(a,\d):=||Au_a-f_\d||=C\d,\,\, C=const\geq 1,\,\, ||f_\d||>\d.
\eqno{(1.4)}
$$
 Note that $N(Q)=N(A^*)$, where  $Q:=AA^*\geq 0,$ and
$$
||Au_a-f_\d||=||[A(B+a)^{-1}A^*-I]f_\d||=||Q(Q+a)^{-1}-I]f_\d||=
a||(Q+a)^{-1}f_\d||,
$$
where the known formula $(B+a)^{-1}A^*=A^*(Q+a)^{-1}$ was used.

Thus,
$$h=a||(Q+a)^{-1}f_\d||.
$$
This implies that, for a fixed pair $\{f_\d, \d\},$ the 
function $h(a,\d)$ in (1.4) is a monotone
increasing function of $a$ on $(0,\infty)$, $\lim_{a\to 0}h(a,\d)=0$,
where we have used the assumption $f_\d\perp N(A^*)$, and 
$h(\infty, \d)=||f_\d||>\d$, so that (1.4) has a unique solution
$a=a(\d)$, $\lim_{\d\to 0}a(\d)=0$.

Define $R_\d f_\d:=u_\d:=u_{a(\d)}$, where $a(\d)$ is given by the DP.

{\bf Theorem 1.1.} {\it If $A$ is compact, i) holds, and $R_\d 
f_\d:=u_\d$, then (1.1) 
holds.
}

{\bf Proof.} From (1.4) and (1.3), with $a=a(\d)$ and $u_a:=u_\d$, one 
gets
$$
||u_\d||\leq ||y||.
\eqno{(1.5)}
$$
Thus, $u_\d\rightharpoonup u$ (weak sequential convergence) as $\d\to 0$,
and $Au_\d \to Au$, because $A$ is compact. From (1.3), with $\d\to 0$ and 
$a\to 0$, it follows that $Au=f$. By the injectivity of $A$, one gets 
$u=y$, so $u_\d\rightharpoonup y$ as $\d\to 0$.
To prove that 
$$
\lim_{\d\to 0}||u_\d-y||=0,
\eqno{(1.6)}
$$
it is sufficient to prove that $\lim_{\d\to 0}||u_\d||=||y||$, and
this follows from (1.5) because
$||y||\leq \liminf_{\d\to 0} ||u_\d||\leq \limsup_{\d\to 0} ||u_\d||\leq 
||y||$. The first
inequality is the lower semicontinuity of the norm in $H$, and  
the last one follows from (1.5). Theorem 1.1 is proved. $\Box$
 
An alternative proof of (1.6) is based on the inequality
$$
||u_\d-y||^2\leq 2\Re (y,y-u_\d),
\eqno{(1.7)}
$$
which follows from (1.5) easily. Since $u_\d\rightharpoonup y$ as $\d\to 
0$, (1.7) implies (1.6). $\Box$

\section{Generalizations. }

2.1. {\it One can drop the injectivity of $A$ assumption
and still get (1.6)}: 

If $N:=N(A)$ is the null-space of 
$A$, $Ay=f$, and $y\perp N$, then $u_\d:=u_{a(\d)}$, defined in (1.2),
has the property $u_\d\perp N$. Indeed, if
$\phi\in N$, then $(u_\d, \phi)=(f_\d, A(B+a)^{-1}\phi)=
(f_\d,(Q+a)^{-1}A\phi)=0$.
Thus $\lim_{\d\to 0} u_\d=u\perp N$. Since the problem
$$
Au=f, \quad u\perp N
\eqno{(2.1)}
$$
has only one solution, it follows that $u=y$. We have proved:

{\bf Theorem 2.1.} {\it If $N\neq \{0\}$ and $y$ solves (2.1), then (1.6) 
holds.
}

2.2. {\it Let us drop both the compactness of $A$ and the injectivity of 
$A$ assumptions.} 

{\bf Theorem 2.2.} {\it If $A$ is a linear bounded operator, and $y$ 
solves (2.1), then (1.6) holds.
}

{\bf Proof.} As above, one gets (1.5) with $u_\d:=u_{a(\d)}$ defined by 
(1.2). Thus,  $u_\d\rightharpoonup u$ and $(B+a)u_\d\to A^*f$ as $\d\to  
0$, and  this implies $Bu=A^*f$. Indeed,
$B\geq 0$ is a monotone
continuous operator, and therefore it is weakly closed, i.e., 
$u_\d\rightharpoonup u$ and 
$Bu_\d\to g$ imply $Bu=g$. Recall that $a=a(\d)\to 0$ and $u_\d$ is 
bounded
as $\d \to 0$, so $\lim_{\d \to 0}a(\d)u_\d=0$. Since $f=Ay$, one has 
$B(u-y)=0$,
and, since $N(B)=N(A)$, one gets $Au=Ay=f$.
We have already proved that $u\perp N$, thus $u=y$, and 
$u_\d\rightharpoonup y$ as $\d\to 0$. This and (1.7) imply (1.6). $\Box$
 
2.3. {\it Let us drop the linearity of $A$ assumption and propose an a 
priori choice of the regularization parameter. }

Consider the functional (cf [3]):
$$
F(u):=||Au-f_\d|| +\d||u||_1, \quad D(F)=H_1,
\eqno{(2.2)}
$$
where $D(F)$ is the domain of $F$, the norm $||\cdot ||_1$ is a norm of a 
Hilbert space $H_1$, which is 
dense in $H$, complete, and the imbedding $i:H_1\to H$ is compact. 
Let  
$$m=m(\d):=\inf_u F(u),$$
and let $u_\d$ be any element such that 
$$
F(u_\d)\leq m+\d.
\eqno{(2.3)}
$$

{\bf Theorem 2.3.} {\it Assume that $A$ is a nonlinear, continuous, 
 injective map,  $A^{-1}$ is not continuous, $f=A(y)$,
$y\in H_1$, and (2.3) holds. Then (1.6) holds.
} 

{\bf Proof.} Let $F(u_n)\to m$. Then for all sufficiently large $n$ one 
has:
$$
m\leq ||A(u_n)-f_\d||+\d ||u_n||_1\leq \d +\d ||y||_1:=c_1\d.
\eqno{(2.4)}
$$
Choose $u_\d:=u_n$ such that (2.3) holds, and let $\d \to 0$.
Since $||u_\d||_1\leq c_1$, one has $u_\d\to u$, $A(u_\d)\to A(u)$, 
and (2.4) shows that $A(u)=f$. By the injectivity of $A$, one gets $u=y$,
so (1.6) holds. $\Box$.

\section{DP does not yield uniform with respect to $f$ convergence. }

In this Section the notations of Section 1 are used.

{\bf Theorem 3.1} {\it Let $S_\d:=\{v: ||Av-f_\d||\leq \d\}$, 
$u_\d:=u_{a(\d)}$, $a(\d)$ solves (1.4), $A$ satisfies assumptions i) of 
Section 1, and $N(A^*)=\{0\}$.

Then there are $f_\d$ such that 
$$
\lim_{\d \to 0} \sup_{v\in S_\d} ||u_\d-v|| \geq c=const>0.
\eqno{(3.1)}
$$
}
{\bf Proof.}
Denote $(B+a)^{-1}A^*:=T$,  $B:=A^*A$, and $u_\d=Tf_\d$.
Then  $||T||=\frac 1 {2\sqrt{a}}$ (see, e.g., [1]). Thus,
there exists a $p=p_a$, such that 
$$||Tp||\geq \frac \d {8\sqrt{a}}, \quad ||p||=0.5\d.$$
Let $R(A)$ denote the range of $A$. Since $R(A)$ and 
$R(B)$ are dense in $H$, and $N(A^*)=\{0\}$, there is a $z=z_a$ such that 
$$
||f_\d-AB^bz-p||\leq \frac {\d} {8}, \quad b=const\in (0,1)\quad 
||p||=0.5\d.
\eqno{(3.2)}
$$
For any $v$, one has:
$$
||Tf_\d-v||\leq ||Tf_\d-TAv||+||TAv-v||.
\eqno{(3.3)}
$$
If $v=B^bz$, then
$$
\lim_{\d \to 0} \sup_{v\in S_\d, \, v=B^bz, ||z||\leq M}||TAv-v||=0,
\eqno{(3.4)}
$$
where $M>0$ is an arbitrary large constant, and we have used the following 
formulas: 
$$TAv-v= [(B+a)^{-1}B-I]B^bz=-a(B+a)^{-1}B^bz,$$
$$||a(B+a)^{-1}B^b||=a\sup_{0\leq s \leq ||B||} \frac 
{s^b}{s+a}=ca^b,\quad
c=b^b (1-b)^{1-b}.$$

From (3.3) and (3.4) it follows that
$$\lim_{\d \to 0} \sup_{v\in S_\d,\, v=B^bz,\, ||z||\leq M} 
||Tf_\d-v||=0
\eqno{(3.5)}
$$ 
if and only if 
$$\lim_{\d \to 0} \sup_{v\in S_\d,\, v=B^bz,\, ||z||\leq M}
||Tf_\d-TAv||=0.
\eqno{(3.6)}
$$  
Choose $v=B^bz$, where $z:=z_a.$ Then, using the triangle
inequality and (3.2), one gets: 
$$||f_\d-Av||\leq ||p||+\frac \d 8 \leq \d,
\eqno{(3.7)}
$$
and
$$||Tf_\d-TAv||=||T(f_\d-Av-p)+Tp||\geq ||Tp||-\frac {\d} {8}||T||\geq 
\frac 
{\d}{\sqrt{a}}(\frac 1 8 
-\frac 1 {16})= \frac \d{16\sqrt{a}}.
\eqno{(3.8)}
$$

If one finds $f_\d$ for which equation (1.4) has a root
$$
a=c_1\d^2[1+o(1)],\quad c_1=const>0, \quad \d\to 0,
\eqno{(3.9)}
$$
then $\frac \d {\sqrt{a}}\geq c=const>0$, and (3.8)
implies that (3.6) fails, so (3.5) fails, and Theorem 3.1 follows.

There are infinitely many such $f_\d$. Let us construct one of them.
Assume, for simplicity, that $A=A^*$, so $B=A^2$, and (1.4) is:
$$
a^2\sum_{j=1}^\infty \frac {g_j}{(\mu_j+a)^2}=C^2\d^2, \quad 
g_j:=|f_{\d j}|^2, \quad \sum_{j=1}^\infty g_j<\infty,
\eqno{(3.10)}
$$
where $\mu_j>0$ are the eigenvalues of $A^2$. Let $g_j=j^{-2},\, 
\mu_j=j^{-1}$. Then, as we prove below,
$$
\phi:=\sum_{j=1}^\infty \frac {j^{-2}}{(j^{-1}+a)^2}= a^{-1}[1+O(a)]
\quad a\to 0.
\eqno{(3.11)}
$$
If (3.11) holds, then (3.10) yields (3.9), and the argument is completed.

To prove (3.11), define $\psi:=\int_1^\infty x^{-2}(x^{-1}+a)^{-2}dx$,
and, setting $t=x^{-1}$, one gets 
$$\psi=\int_0^1(t+a)^{-2}dt=a^{-1}[1+O(a)] \quad a\to 0.$$
This
and the estimate 
$$\phi -(1+a)^{-2}<\psi <\phi,$$
 yield (3.11). $\Box$

In our construction, $w:=f_\d\notin R(A)$ does not depend on $\d$, and
$f=Av_\d$ is chosen for each $\d>0$ in a $\d-$neighborhood of $f_\d$
(see (3.7)),
so that $f$ depends on $\d$. The element $v_\d=B^bz_a,\, a=a(\d),$ cannot 
converge as $\d\to 0$: if $v_\d\to v$ and $Av_\d \to w$ then,
by the continuity of $A$, one has $Av=w$. This is a contradiction, since 
$w\notin R(A)$. This contradiction proves that $v_\d$ cannot converge.
If $A$ is compact, then a similar argument proves that $\lim_{\d\to 
0}||v_\d||=\infty$.

\end{document}